\documentclass{aa}  
\usepackage{graphicx}
\usepackage{txfonts}
\usepackage{psfig,longtable,lscape}
\usepackage{epsfig}
\usepackage{graphicx}
\usepackage{txfonts}
\usepackage[]{natbib}
\bibpunct{(}{)}{;}{a}{}{,} 
\def \inte {{$INTEGRAL$}}
\def \xmm {{\em XMM--Newton}}

\def \sw {{\em Swift}}
\def \src {\mbox{IGR~J11215--5952}}
\def \degmark{^\circ}

\def \hcm {\hbox {\ifmmode $ atom cm$^{-2}\else atom cm$^{-2}$\fi}}

\def \arcsec {\hbox{$^{\prime\prime}$}}
\def \chisq {$\chi ^{2}$}

\def \mdot {\dot{M}_{w}}

\defcitealias{SidoliPM2006}{Paper~I}
\defcitealias{Romano2007}{Paper~II}

\begin{document}
   \title{An alternative hypothesis for the outburst mechanism in \\ 
Supergiant Fast X--ray Transients: the case of IGR J11215--5952}

   \author{L.\ Sidoli\inst{1}, P. Romano\inst{2,3}, S.\ Mereghetti\inst{1}, 
A.\ Paizis\inst{1}, S.\ Vercellone\inst{1}, V.\ Mangano\inst{4} 
	 \and  D.\ G\"{o}tz\inst{5} }
   \offprints{L.\ Sidoli, sidoli@iasf-milano.inaf.it}
   \institute{INAF, Istituto di Astrofisica Spaziale e Fisica Cosmica, 
	Via E.\ Bassini 15,   I-20133 Milano,  Italy
    \and INAF, Osservatorio Astronomico di Brera, Via E.\ Bianchi 46, 
	I-23807 Merate (LC), Italy 
    \and Universit\`a{} degli Studi di Milano, Bicocca, Piazza delle Scienze 3, 
	I-20126 Milano, Italy  
    \and INAF, Istituto di Astrofisica Spaziale e Fisica Cosmica, 
	Via U.\ La Malfa 153, I-90146 Palermo, Italy 
    \and CEA-Saclay, DAPNIA/Service d'Astrophysique, F-91191 Gif-sur-Yvette Cedex, France. 
             }
   \date{Received 22 June 2007/Accepted 3 October 2007}
\abstract
{The physical mechanism responsible for the short outbursts in a recently recognized 
class of High Mass X--ray Binaries, the Supergiant Fast X--ray Transients (SFXTs), is still unknown.
Two main hypotheses have been proposed to date: the sudden accretion by the compact object 
of small ejections originating in  a clumpy wind
from the supergiant donor, or outbursts produced at (or near) the periastron passage in 
wide and eccentric orbits, in order to explain 
the low ($\sim$10$^{32}$~erg~s$^{-1}$)  quiescent emission.
Neither  proposed mechanisms seem to explain the whole phenomenology of these
sources. 
}
   {Here we propose a new explanation for the outburst mechanism, based on the X--ray observations of 
the unique SFXT known to display periodic outbursts, IGR~J11215-5952.
}
{We performed three Target of Opportunity observations with \sw, \xmm\ and \inte\
at the time of the fifth  outburst, expected on 2007 February 9. 
\sw\ observations of the February 2007 outburst have been reported elsewhere. 
Another ToO with \sw\ was performed in July 2007,
in order to monitor the supposed ``apastron'' passage.
}
{\xmm\ observed the source on 2007 February 9, for 23~ks, at the peak of the
outburst, while \inte\ started the observation two days later, failing to detect the source,
which had already undergone the decaying phase of the fast outburst.
The Swift campaign performed in July 2007 reveals a second outburst occurring on 2007 July 24,
as bright as that observed about 165~days before.
}
{The new X--ray observations 
allow us to propose an alternative hypothesis for the outburst mechanism in SFXTs, linked to
the possible presence of a second wind component, 
in the form of an equatorial disk from the supergiant donor. 
We discuss the applicability of the model to 
the short outburst durations of all other Supergiant Fast X--ray Transients, 
where a clear periodicity in the outbursts has not been found yet.
The new outburst from \src\ observed in July suggests 
that the true orbital period is  $\sim$165~days,
instead of 329~days, as previously thought.
}
\keywords{X-rays: binaries - stars: neutron - accretion - X-rays: stars: individual: \src  }
\authorrunning {L.\ Sidoli et al.}
\titlerunning {An alternative model for the outburst from Supergiant Fast X--ray Transients}

\maketitle

\section{Introduction}

The Galactic plane monitoring performed by the \inte\ satellite
in the last 5 years has allowed the discovery of a number of 
new High Mass X--ray Binaries (HMXBs).
Several of these new sources are intrinsically highly absorbed
and were difficult
to discover with previous missions (e.g. IGR~J16318--4848, \citealt{Walter2003}).
Others are transient HMXBs (associated with OB supergiant) 
displaying short outbursts 
(few hours, typically less than a day; 
\citealt{Sguera2005}), and form the recently recognized new
class of  Supergiant Fast X--ray Transients (SFXTs).

\object{IGR~J11215--5952} is a hard X--ray transient  discovered by 
\inte\  during a fast outburst in  April 2005 \citep{Lubinski2005}.
The short duration of this outburst 
led \citet{Negueruela2005a} to propose that 
\src\ could be a new member of the 
class of Supergiant Fast X-ray Transients.
The optical counterpart is indeed a B-type supergiant,
\object{HD~306414} located at a distance of 6.2~kpc 
(\citealt{Negueruela2005b}, \citealt{Masetti2006}, \citealt{Steeghs2006}).

From the analysis of archival \inte\ observations and the discovery
of two previously unnoticed outbursts, 
a recurrence period in the X--ray activity of $\sim$330~days
has been found \citep*[hereafter Paper I]{SidoliPM2006}, likely
linked to the orbital period of the binary system.
This periodicity was later confirmed by the fourth outburst
from \src\ observed with $RXTE/PCA$  on 2006 March 16--17, 
329~days after the previous one \citep{Smith2006a}. 
The \inte\ spectrum was well fitted by a hard power-law 
with a high energy cut-off around 15~keV (Paper~I).
Assuming a  distance of 6.2 kpc, the peak fluxes of the outbursts
correspond to a luminosity of $\sim 3 \times$10$^{36}$~erg~s$^{-1}$ (5--100~keV; Paper~I).
The $RXTE/PCA$ observations showed  
a possible pulse period of  $\sim$195\,s \citep{Smith2006b},
later confirmed during the February 2007 outburst,
yielding P=$186.78\pm0.3$\,s \citep{Swank2007}.
 
All these findings confirmed \src\ as a member of the 
class of SFXTs, and  the first one displaying periodic outbursts. 
Based on the known periodicity, an outburst was expected for 2007 February 9. 
This  allowed us to obtain several
Target of Opportunity (ToO) observations  with $Swift/XRT$,  $XMM-Newton$ 
and $INTEGRAL$.
The $Swift/XRT$ results of the February 2007 outburst 
have been reported in \citealt{Romano2007} 
(hereafter Paper~II).
Here we report the results of the \xmm\ and \inte\ observations 
of the February 2007 outburst, 
and of a $Swift/XRT$ campaing performed in July 2007, in order to monitor
the supposed apastron passage.

  \section{Observations and Data Reduction}

\subsection{\xmm\ data}

\begin{figure}
\vbox{
\includegraphics[height=9.cm,angle=-90]{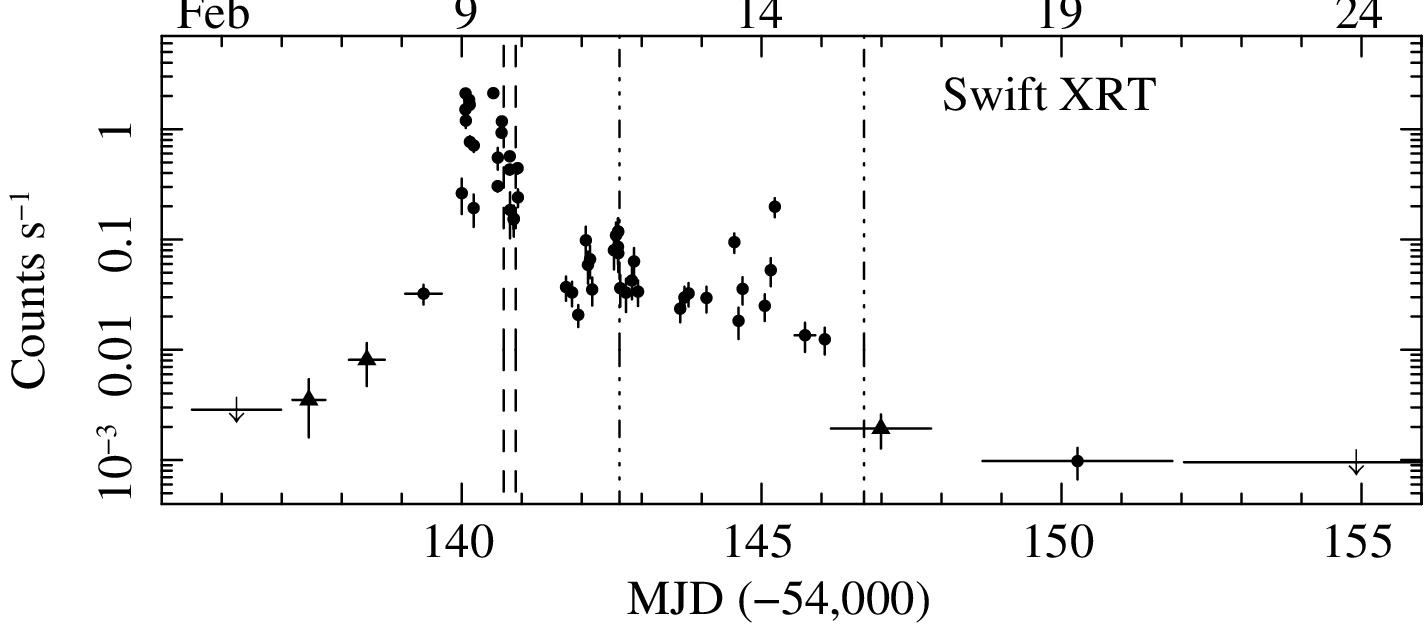}
\includegraphics[height=9.cm,angle=-90]{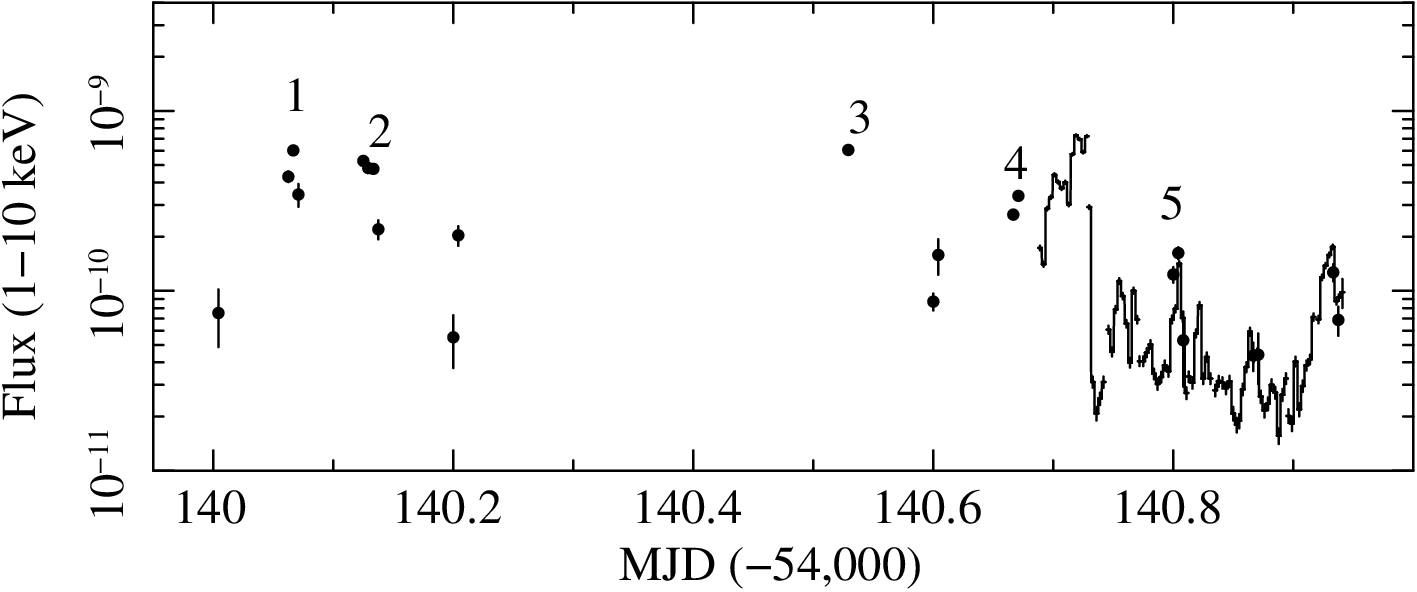}}
\caption[]{{\em Upper panel:} $Swift/XRT$ lightcurve of the February 2007 outburst from \src\ (from Paper~II)
together with the start and end times 
of the \xmm\ (dashed vertical lines) and of the \inte\ (dashed-dotted vertical lines) 
observations (see Table~\ref{tab:log} for details). Day 140 corresponds to 2007 February 9.
{\em Lower panel:} Close-up view of the brightest part of the \src\ lightcurve observed with $Swift/XRT$ 
(solid circles, 1--10 keV), on 2007 February 9 (adapted from Paper~II) 
together with the \xmm\ lightcurve (solid line, PN data, 1--10~keV).
The flux is corrected for the absorption and 
in units of erg~cm$^{-2}$~s$^{-1}$ in the 1--10~keV range, obtained assuming a power-law spectrum with photon
index of 1 and  a column density of 10$^{22}$~cm$^{-2}$ (the numbers mark the ``flares'' named as
in Paper~II).
}
\label{fig:comp}
\end{figure}

\src\ was observed with 
XMM-Newton  on 2007 February 9, with a net exposure
of 23.3 ks (see Table~\ref{tab:log} for the observations log).
The observation covered
a small fraction of 
the brightest part of the outburst. In Fig.~\ref{fig:comp} (upper panel) 
we show the times
of the \xmm\ and \inte\ observations compared with the
$Swift/XRT$ simultaneous lightcurve, reported in Paper~II. 

Data were reprocessed using version 6.5 of the Science Analysis
Software (SAS). Known hot, or flickering, pixels and electronic
noise were rejected. 
The response and ancillary matrices
were generated using the SAS.
EPIC PN operated in Small Window mode, while both MOS cameras were in Full Frame mode.
MOS and PN observations used the medium thickness filter.
Spectra were selected 
using pattern from 0 to 4 with the PN, and from 0 to 12 with both MOS cameras.
Source counts were extracted from circular regions
of 40\arcsec\ radius for the PN and 1$'$ for MOS1 and MOS2.
With the SAS task {\em epatplot} we verified that both PN and MOS data
were not affected by pile-up.
Background counts were obtained from similar
regions offset from the source position. The background 
showed evidence for a moderate flaring activity only in the PN, almost
at the end of the observation, so we further selected PN in order to exclude
this small portion of the observation. No further time selection has been applied
to data from both MOS cameras.

To ensure applicability of the \chisq\ statistics, the
net spectra were rebinned such that at least 20 counts per
bin were present and such that the energy resolution was not
over-sampled by more than a factor 3.  All spectral
uncertainties and upper-limits are given at 90\% confidence for
one interesting parameter.

\subsection{INTEGRAL data}

The \inte\ observations were performed about 2 days 
after the \xmm\ observations, i.e. from 
2007 February 11 to February 15, and consist of 
a total of 101 pointings ($\sim$2900\,s each, see Table~\ref{tab:log} for details).
Version 6.0 of the Off-line Scientific Analysis (OSA) software
has been used to analyse the data. 
For each pointing, we extracted images in the \mbox{22--60} and 
\mbox{60--100\,keV} energy bands for IBIS/ISGRI 
and 3--10 and 10--25\,keV bands for JEM--X1.
For both instruments, the images were used to build a mosaic sky map.
Single pointing as well as mosaic results were inspected.

\begin{table}[h!]
\caption{Summary of the \xmm\ and \inte\ pointed observations of \src. 
} 
\label{tab:log}
\begin{tabular}[c]{ccc}
\hline\noalign{\smallskip}
\hline\noalign{\smallskip}
  Start time     & End time     & Net Exp.  \\
 (dy~mon~yr~hr:mn:ss)     & (dy~mon~yr~hr:mn:ss)     & (ks)         \\
\noalign{\smallskip\hrule\smallskip}
\xmm:    &   &   \\
  9 Feb 2007 16:05:56 & 9 Feb 2007 22:34:16    & 23.3   \\
\noalign{\smallskip\hrule\smallskip}
\inte:     &   &   \\
 11 Feb 2007 15:07:58 &   12 Feb 2007 23:12:49  &  111.0 \\
       12 Feb 2007 23:40:07 &   13 Feb 2007 11:46:59  &  42.0 \\
       13 Feb 2007 12:13:34 &   13 Feb 2007 19:08:24  &  24.0 \\
       13 Feb 2007 19:41:00 &   14 Feb 2007 05:11:37  &  33.0 \\
       14 Feb 2007 14:58:35 &   15 Feb 2007 16:58:23  &  90.0 \\
\noalign{\smallskip\hrule\smallskip}
\end{tabular}
\end{table}


\subsection{\sw\ data}

\sw\ observations of the February 2007 outburst have been already discussed in Paper~II.
The $Swift/XRT$ data reported here were obtained as a ToO 
campaign aimed at studying the source close to apastron, expected on 2007 July 24. 
Since \src\ at apastron would have been $<4$ hours from the Sun,
hence unobservable by Swift under normal circumstances, we spread the 
$\sim$15\,ks estimated to be required for a detection (or firm upper limit)
during quiescence over a roughly two-month time, in short (1.5--2.5\,ks) 
integration times. 
As a result (see Table~\ref{igr112:tab:alldata2}), 
\src\ was observed starting on 2007-06-05 17:52 UT 
through 2007-07-31 10:27 UT, for a total on-source exposure of 16.8\,ks. 

The XRT data were processed with the same standard procedures as 
described in Paper~II (using the {\tt Heasoft} package, v.6.3.1;
spectral redistribution matrices v010).

  \section{Analysis and Results}

\subsection{February 2007 outburst}

	\begin{figure*}[th!]
\includegraphics[angle=-90,width=16cm]{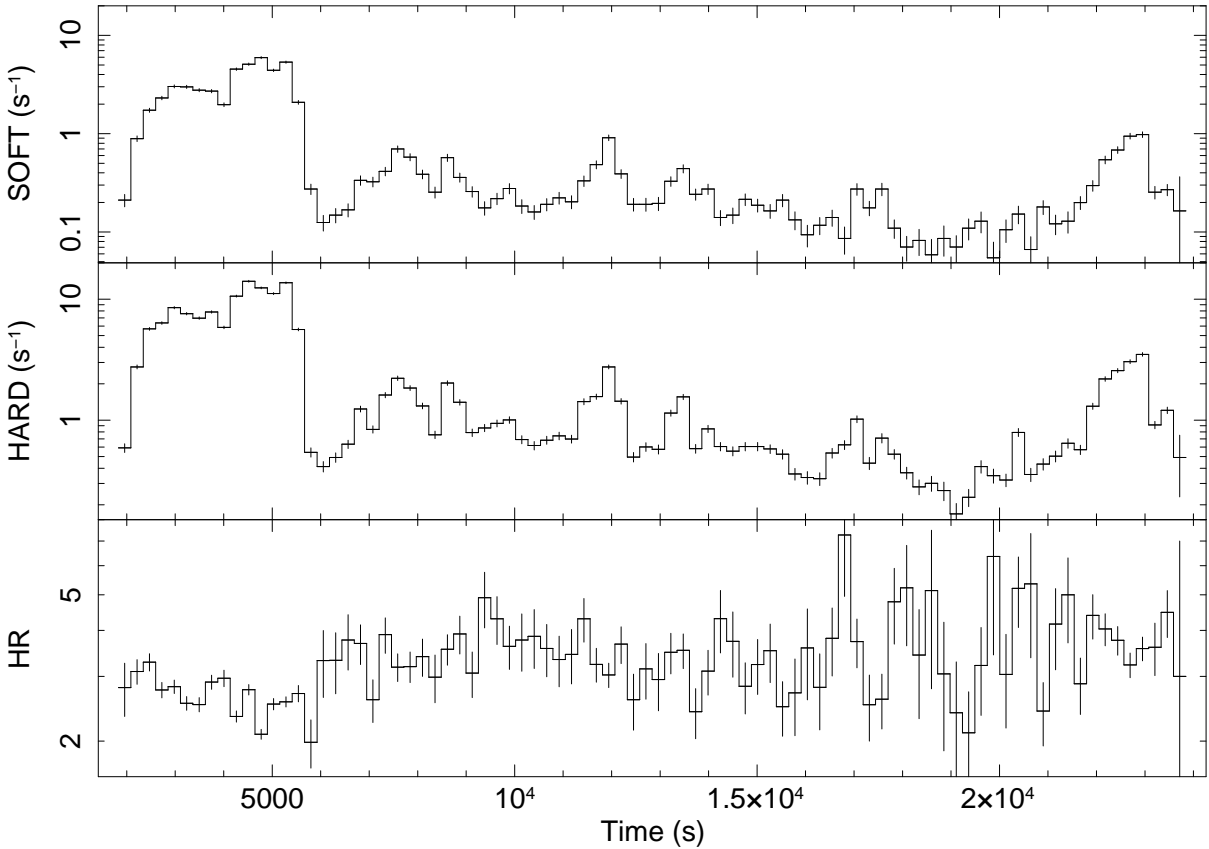}
 		\caption{\src\ background subtracted 
lightcurve observed with the PN camera, in two energy bands 
(upper panel, 0.3--2 keV; middle panel, 2--12~keV) together with their hardness ratio (lower panel).
A large variability is evident, with a ``bright flare'' 
at the beginning of the observation (lasting $\sim$`~hr), followed by a ``faint'' 
phase with a lower count rate level. 
During this ``faint'' level of X--ray emission, smaller level flaring activity is also present. 
Time is in seconds from the beginning of the observation. Bin time is 256~s.		}
                \label{fig:netlc}
	\end{figure*}

The \src\ lightcurve of the brightest part of the February 2007 outburst, obtained by combining
the $Swift/XRT$ and the  \xmm\ PN data, is shown in Fig.~\ref{fig:comp} (lower panel). 
The PN data are shown in more detail in Fig.~\ref{fig:netlc}, where the lightcurves in the soft (0.3--2~keV)
and hard (2--12~keV) bands and their ratio are displayed.
A bright flare is present at the beginning of 
the observation (which was not caught with $Swift/XRT$), 
lasting about 1~hour,
and displaying large variability (hereafter part ``A'' of the \xmm\ observation, from the beginning up to 5500~s
in Fig.~\ref{fig:netlc}). 
Then a lower intensity activity follows (hereafter part ``B'' of the observation, from about 6000~s to the end of the
observation), 
also composed by several lower intensity flares.

The hardness ratio as a function of time seems to display some level of variability: 
fitting it to a constant model yields a 
reduced $\chi^2=3.95$ for 85 degrees of freedom (dof). 
Fitting to a constant the ``A'' and ``B'' parts we obtain
a value of $2.5$ and reduced $\chi^2=8.22$ (for 13 dof) for the ``A'' part, and 
a value of $3.3$ and reduced $\chi^2=1.22$ (for 69 dof) for the ``B'' part.

\begin{figure}[th!]
\vbox{
\includegraphics[height=8.cm,angle=-90]{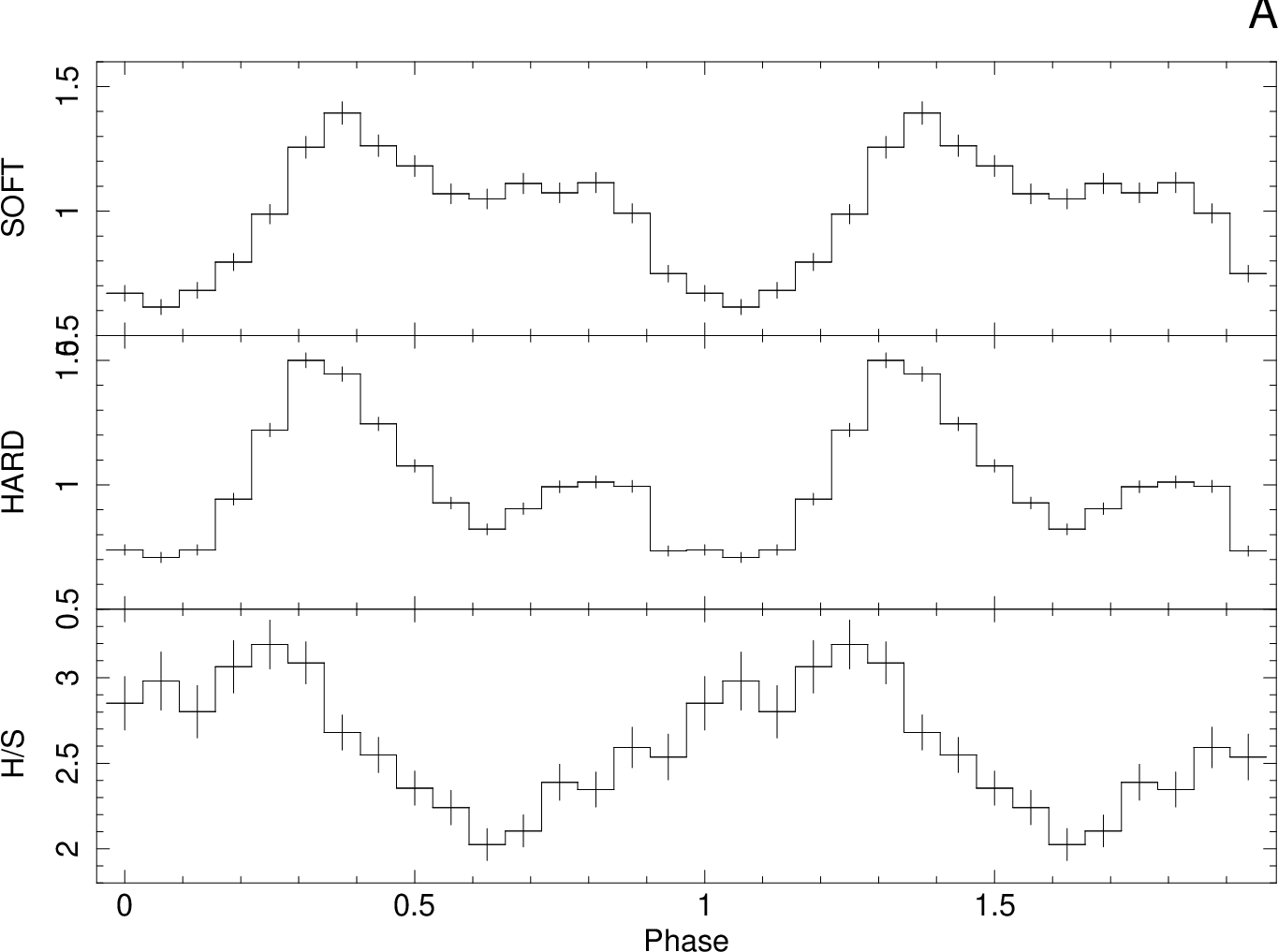}
\includegraphics[height=8.cm,angle=-90]{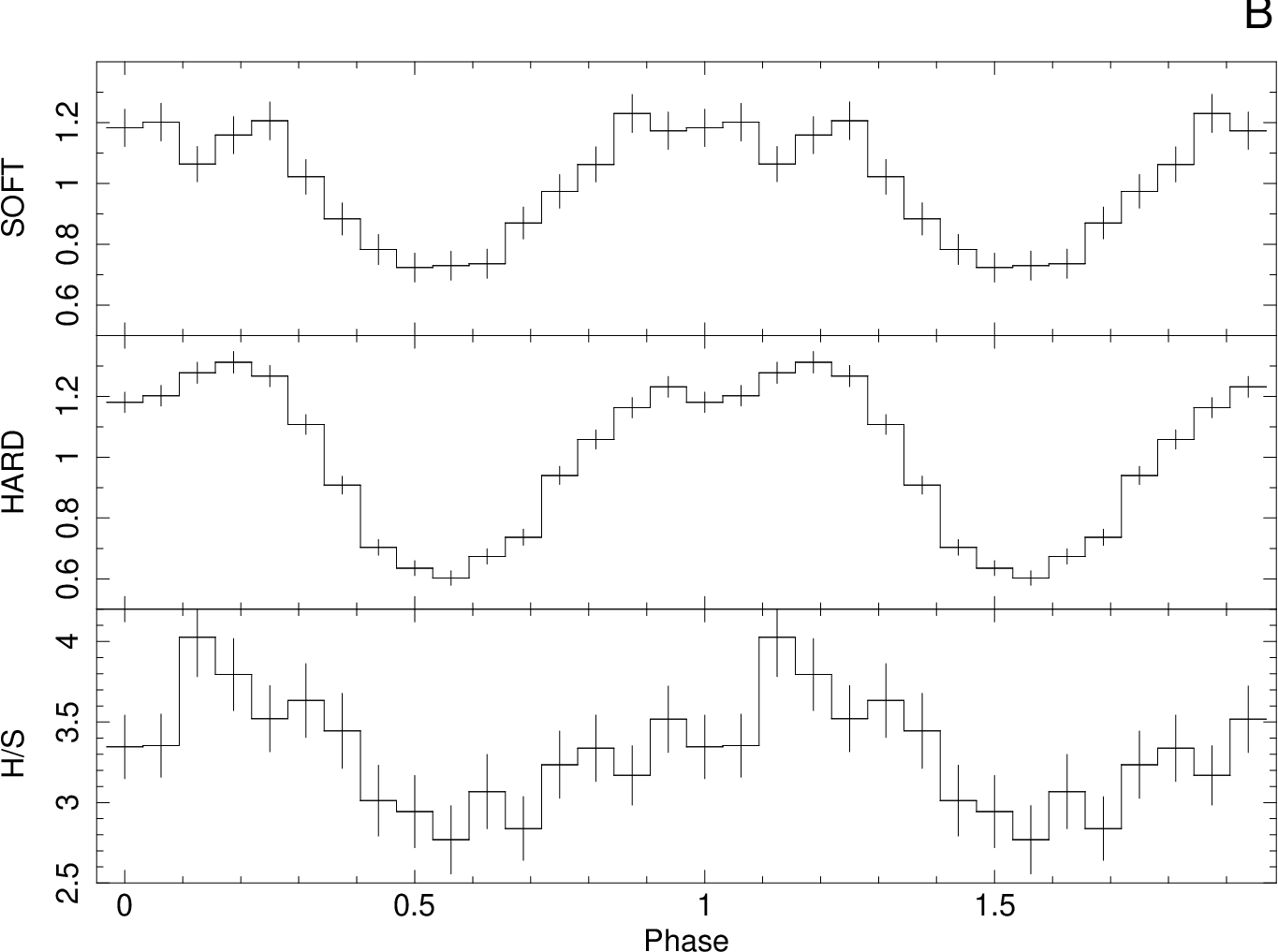}}
\caption[]{\src\ lightcurves (normalized intensities)
in different energy bands (``soft'' means 0.3--2 keV, ``hard'' 2--12~keV),
folded on the period of 187~s, together with their hardness ratios H/S. 
From  top to bottom, ``A'' is the folded lightcurve of the ``bright'' part,
while ``B'' is the folded lightcurve of the ``faint'' second part of the observation.
Note that all lightcurves assume the same starting epoch (MJD=54140).
}
\label{fig:efold}
\end{figure}

After correcting the arrival times  to the Solar System barycenter,
we searched for the \src\ spin periodicity by using epoch 
folding techniques,
finding a period of 186.94$\pm0.58$\,s, compatible with that derived
by $XTE/PCA$ data \citep{Swank2007}.

The folded lightcurves (EPIC PN, background subtracted) 
in two different energy ranges 
are shown in
Fig.~\ref{fig:efold}, together with their hardness ratios.
The pulse profile changes between the ``bright'' 
and the ``faint''  part of the observation.
In the bright part (``A'') the folded lightcurve shows a double-peaked shape, and the main
pulse peaks around phase 0.3--0.4 (and the secondary peak at 0.8), while 
the ``B'' lightcurve shows a single broad pulse, around  phase 0.0 (where the ``A''
lightcurve shows a minimum). 

\begin{figure*}[th!]
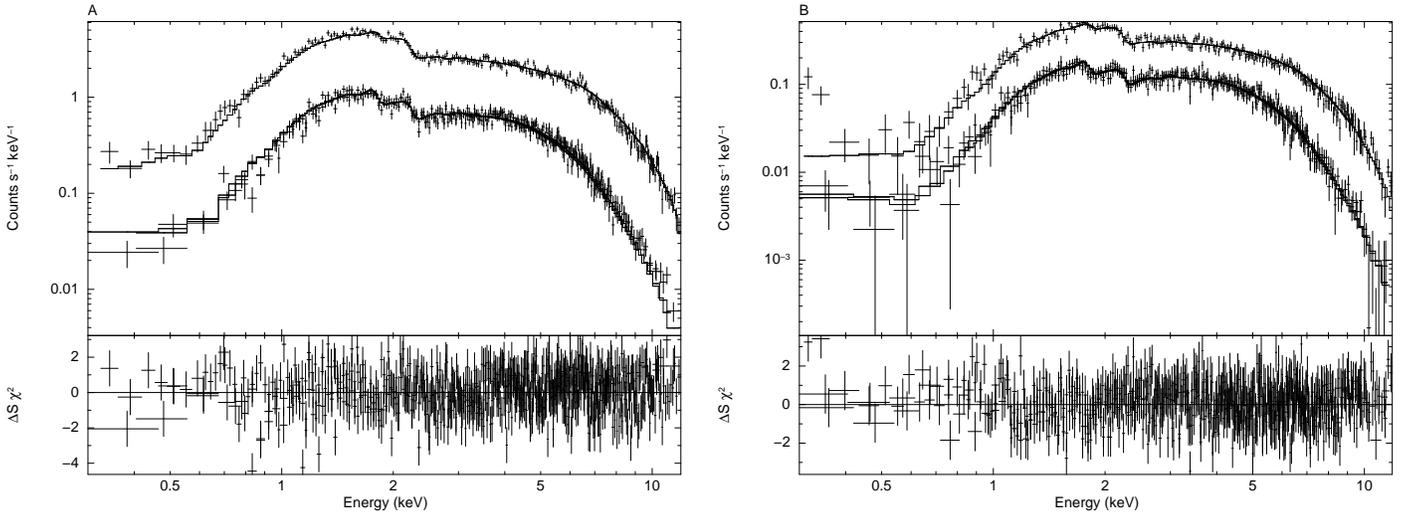

\hbox{
\includegraphics[height=9.cm,angle=-90]{8137fig4a.ps}
\hspace{0.3cm}
\includegraphics[height=9.cm,angle=-90]{8137fig4b.ps}}
\caption[]{\src\ count spectra fitted with an absorbed cut-off powerlaw (see text
for the parameters): in the {\em left panel} the spectra (PN+MOS1+MOS2) 
extracted from the bright initial flare (``A''), while in the 
{\em right panel} the spectra taken from the fainter part of the observation (``B'') are shown,
together with the residuals (lower panels). 
}
\label{fig:spec}
\end{figure*}

\begin{figure}[th!]
                \includegraphics[angle=-90,width=9cm]{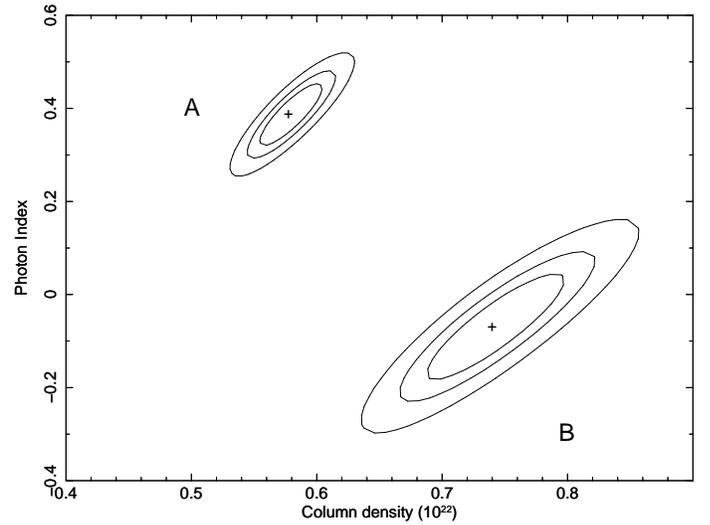}
 		\caption{Confidence contours levels (68\%, 90\% and 99\%) for the
two parameters of the cut-off power-law model applied to
the two intensity selected spectra from the overall ``bright'' (A) and the ``faint'' (B) phases.
		}
                \label{fig:contall}
	\end{figure}

We initially extracted two specta, one from the part ``A'' 
(bright initial flare), and the second from the part ``B'' (the fainter second part).
Fitting together PN, MOS1 and MOS2 spectra, an absorbed power-law does not fit the data well. 
Good results were
obtained with a cut-off power-law, giving an
absorbing column density of ($0.57\pm{0.03})\times$10$^{22}$~cm$^{-2}$ (``A'')
and  ($0.7 \pm{0.1})\times$10$^{22}$~cm$^{-2}$  (``B''), a photon index $\Gamma$=0.37$\pm{0.09}$  (``A'')
and $\Gamma$=0.0$\pm{0.16}$  (``B''), a high energy cut-off, of 7$\pm{1}$~keV  (``A'') and
of 4$\pm{1}$~keV
(``B''; see Fig.~\ref{fig:spec} for the two 
count spectra and Fig.~\ref{fig:contall} for the confidence contour plots). 

\begin{figure}[t]
                \includegraphics[angle=-90,width=9cm]{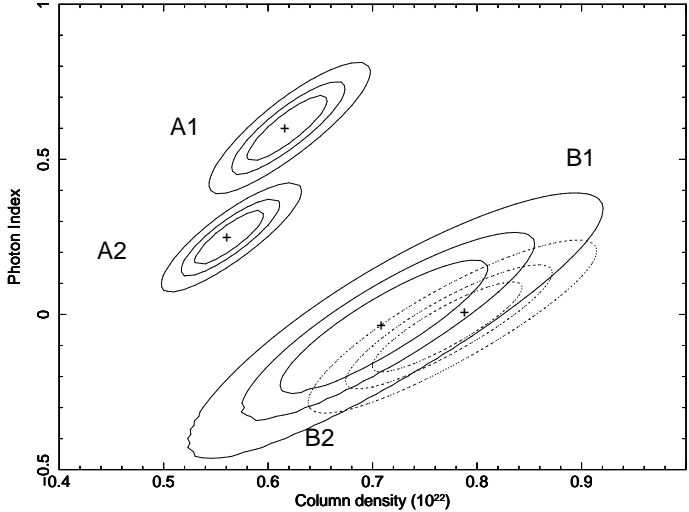}
 		\caption{Confidence contours levels (68\%, 90\% and 99\%) for the
two parameters of the cut-off power-law model applied to
the four spin-phase selected spectra reported in Table~\ref{tab:phasespec}.
Dotted lines refers to the B2 spectrum.
		}
                \label{fig:cont}
	\end{figure}

From the folded lightcurves (Fig.~\ref{fig:efold}), 
there is evidence that the hardness ratios vary
with the spin phase, with lower hardness ratios concentrated in the
pulse phase [0.4--0.8] (independent from the source state, ``A'' or ``B'').
This suggested to perform a phase resolved spectroscopy,
extracting spectra
for each of the two source intensity states, in two phase intervals: [0.4--0.8]  (phase resolved spectra ``1'',
A1 or B1, see Fig.~\ref{fig:cont})
and (0.8--1.4) (phase resolved spectra ``2'',
A2 or B2).
It is also evident from the two hardness ratios shown in Fig.~\ref{fig:efold} 
that when the source is brighter (state ``A''), it is also softer (on average). 


An absorbed power-law spectrum is never the best-fit to all the four analysed spectra,
leading to structured residuals, particularly evident when fitting the spectra
extracted from spin phase (0.8--1.4).
This suggested the addition of a soft component (a blackbody), or the use of a power-law with
a high energy cut-off.
The fit results are equally adequate in the two cases (see Table~\ref{tab:phasespec}).

An emission  line from  iron is never required by the data. 
If we add  a Gaussian line at 6.4~keV
to the cut-off power-law continua of the ``bright'' spectra,
we obtain  only a marginal ($\sim$ 3~$\sigma$ level) detection with
a line flux of (5.7$^{+1.7} _{-1.8}$)$\times$$10^{-5}$~photons~cm$^{-2}$~s$^{-1}$
for phase range [0.4--0.8], and (5.0$^{+2.6} _{-1.6}$)$\times$$10^{-5}$~photons~cm$^{-2}$~s$^{-1}$
for phase range (0.8--1.4) (uncertainties in the total line fluxes are 1~$\sigma$).

\begin{table*}
\caption{Spin-phase spectroscopy of \src\ (PN+MOS1+MOS2 data), with two models:
an absorbed cut-off power-law {\sc cutoffpl} model in {\sc xspec}, and an absorbed power-law together
with a blackbody.
``A'' and ``B'' spectra refer to the ``bright'' and ``faint'' source states, respectively,
while ``1'' and ``2'' mean the two spin phase ranges  [0.4--0.8] and (0.8--1.4), respectively.
$\Gamma$ is the power-law photon index, E$_{\rm c}$ the high energy cut-off
in the {\sc cutoffpl} model,  kT$_{\rm bb}$ and  R$_{\rm bb}$ are the temperature and the
radius of the blackbody emitter (assuming a distance of 6.2~kpc); fluxes are corrected for the absorption,
in the energy range 0.5--10 keV.
} 
\label{tab:phasespec}
\begin{tabular}[c]{cccccccc}
\hline\noalign{\smallskip}
\hline\noalign{\smallskip}
Spectrum   &  N$_{\rm H}$             &  $\Gamma$  & E$_{\rm c}$   &  kT$_{\rm bb}$ & R$_{\rm bb}$     & Flux  &  red. $\chi^{2}$ (dof) \\
           &  (10$^{22}$~cm$^{-2}$)   &            &  (keV)        &  (keV)         &   (km)           &  (10$^{-11}$~erg~cm$^{-2}$~s$^{-1}$)     &  \\
\noalign{\smallskip\hrule\smallskip}
Cut-off powerlaw  &     &     &     &  &    &  &    \\
\noalign{\smallskip\hrule\smallskip}
   A1          &   0.62 $\pm{0.04}$    &     0.60 $\pm{0.11}$  &  11  $^{+4}   _{-2}$      &  $-$ &  $-$   & 6.7 &  1.29 (532)   \\
   A2          &   0.56 $\pm{0.04}$    &     0.25 $\pm{0.10}$  &  7.3 $^{+1.3} _{-1.0}$    &  $-$ &  $-$   & 9.9 &  1.18 (589)    \\
   B1          &   0.71 $\pm{0.11}$    &     0.0 $\pm{0.23}$   &  4.1 $^{+1.1} _{-0.7}$    &  $-$ &  $-$   & 0.5 &  0.99 (427)   \\
   B2          &   0.79 $\pm{0.07}$    &     0.0 $\pm{0.15}$    &  4.8 $^{+0.8} _{-0.6}$    &  $-$ &  $-$   & 1.3 &  1.08 (574)    \\
\noalign{\smallskip\hrule\smallskip}
Powerlaw plus blackbody  &     &     &     &  &    &  &    \\
\noalign{\smallskip\hrule\smallskip}
   A1        & 0.73 $^{+0.07} _{-0.06}$ & 1.23 $^{+0.22} _{-0.16}$ &   $-$  &  2.0 $^{+0.2} _{-0.3}$  &  0.24 $\pm{0.03}$ & 6.7 &  1.29 (531)   \\
   A2        & 0.64 $\pm{0.05}$         & 0.89 $^{+0.11} _{-0.09}$ &   $-$  &  1.7 $\pm{0.2}$         &  0.37 $\pm{0.05}$ & 10  &  1.19 (588)    \\
   B1        & 0.8  $^{+0.1} _{-0.2}$   & 0.96 $^{+0.21} _{-0.39}$ &   $-$  &  1.4 $\pm{0.2}$         &  0.14 $^{+0.06} _{-0.01}$ & 0.5 &  1.0 (426)   \\
   B2        & 0.65 $ ^{+0.11} _{-0.07}$  & 0.4 $\pm{0.3}$    &   $-$  &  1.3 $\pm{0.1}$         &  0.27 $\pm{0.04}$         & 1.3 &  0.99 (573)    \\
\noalign{\smallskip\hrule\smallskip}
\end{tabular}
\end{table*}

Source spectra observed with the RGS arrays (0.3--2.0 keV) resulted into count rates of 
(2.02$\pm{0.15}$)$\times$10$^{-2}$~s$^{-1}$  and 
(1.62$\pm{0.15}$)$\times$10$^{-2}$~s$^{-1}$, respectively with RGS1 and RGS2 (22~ks exposure).
Since they
did not show any evidence for strong emission or absorption lines, 
we will not discuss them further.

The source is not detected by \textit{INTEGRAL} either at the single 
pointing level or in the mosaic. 
We obtain 3$\sigma$ upper limits of $\sim$1\,mCrab and $\sim$5\,mCrab in 
the 22--60 and 60--100\,keV IBIS/ISGRI bands 
($\sim$290\,ks) and upper limits of $\sim$3\,mCrab and $\sim$6\,mCrab in the 
3--10\,keV and 10--25\,keV JEM--X1 bands ($\sim$55\,ks). 

These upper limits can be compared with the simultaneous $Swift/XRT$ observations (see Fig.~\ref{fig:comp},
upper panel). 
The $Swift/XRT$ spectrum 
resulted in a net exposure time of 18~ks. 
Fitting it with an absorbed power-law, we obtain  
an absorbing column density of (1.8$\pm{0.7}$)$\times$10$^{22}$~cm$^{-2}$ with a photon index
of 1.9$\pm{0.5}$. The resulting 3--10~keV flux is 1.4$\times$10$^{-12}$~erg\,cm$^{-2}$~s$^{-1}$, more than
one order of magnitude below the upper limits derived from JEM--X.
The extrapolation of this power-law model to the IBIS/ISGRI 
energy range results in the following
fluxes: 
1.5$\times$10$^{-12}$~erg\,cm$^{-2}$~s$^{-1}$ (22--60~keV) and 
8.6 $\times$10$^{-13}$~erg\,cm$^{-2}$~s$^{-1}$ (60--100~keV), which are again 
more than
one order of magnitude lower than the IBIS/ISGRI upper limits.


\subsection{July 2007 outburst}

A new outburst was observed with $Swift/XRT$ in July 2007, during the
times of the apastron passage, expected based on the orbital period of 329~days.

The source was not detected in the first seven observations (2007 June 5 to July 17),
for which a 3-$\sigma$ upper limit can be placed at $3\times10^{-13}$ erg cm$^{-2}$ s$^{-1}$. 
On July 24, the source flux increased above count rates for which 
pile-up correction is required. Therefore we extracted the source events 
from an annular region with radii 4 and 30 pixels during the
first orbit of data, when the source was piled-up, and from a circular region with a radius 
of 11 pixels for the rest of the observation.
To account for the background, we also extracted events within an 
annular region centred on the source and with radii 40 and 100 pixels, 
free from background sources.

\begin{figure*}
\includegraphics[height=16.9cm,angle=270]{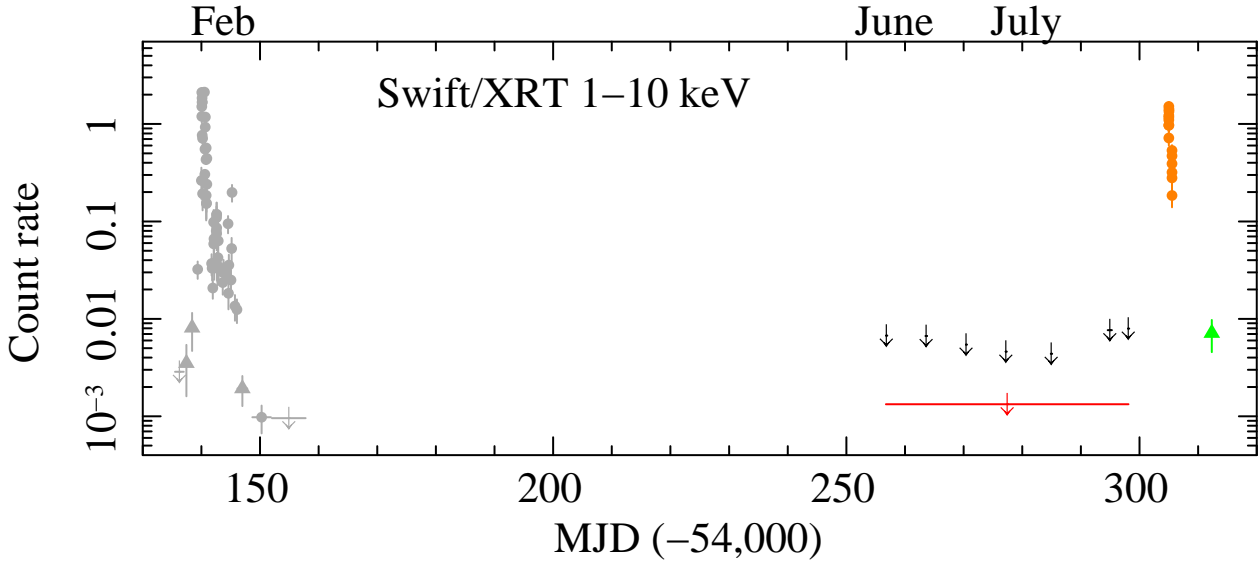}
		\caption{XRT 1--10\,keV light curve, corrected for pile-up, PSF losses, 
		and vignetting and background-subtracted. 
		The grey points were collected during the 2007 February observing campaign (Paper~II).
		The black downward-pointing arrows are 3-$\sigma$ upper limits on single XRT observations 
		(see Table~\ref{igr112:tab:alldata2}) collected between 2007 June 5 and 
		July 17, while the red arrow is the 3-$\sigma$ upper limit obtained combining them.
		The remainder of the data (orange and green points) were collected on 2007 July 24 an 31.
		Different colours denote different observations; 
		filled circles are full detections (S/N$>$3), triangles marginal detections ($2<$S/N$<3$).
		}
 		\label{igr112:fig:lcv2}
	\end{figure*}

Figure~\ref{igr112:fig:lcv2} shows the background subtracted 
1--10\,keV light curve, which is corrected for point-spread function 
(PSF) losses, due to the extraction region geometry, 
bad/hot pixels, and columns falling within this region, and for vignetting.

The spectrum of the  outburst  (observation 31)  was fit 
in the 0.5--9\,keV energy range, adopting a 
single power law, with photon index $\Gamma=0.98_{-0.26}^{+0.28}$, 
and column density of 
$N_{\rm H}=(1.15_{-0.32}^{+0.42})\times 10^{22}$ cm$^{-2}$ 
($\chi^2_{\rm red}=1.08/27$ dof). 
These results are fully consistent with the ones obtained for the 
2007 Feb 9 data (Paper~II). 
The unabsorbed 1--10\,keV flux was $\sim 1.1 \times 10^{-10}$ erg cm$^{-2}$ s$^{-1}$.

\section{Discussion}
\label{discussion}

The observations reported here, together with the $Swift/XRT$ monitoring (Paper~II)
represent the most complete observation of a SFXT outburst, and demonstrate
that the accretion phase in these short transients lasts longer than previously
thought on the basis of less sensitive instruments.
Indeed, only the brightest part of the outburst, which  
is short and lasts a few hours,
would have been seen by the INTEGRAL instruments. 
This is clearly demonstrated by the fact
that actually INTEGRAL, starting the observations only 2 days after 
the onset of the outburst, did not detect the source.

The \xmm\ observation was performed during the bright part of the \src\ 
outburst (on February 9), and shows a large 
intensity variability, which already
emerged during the $Swift/XRT$ observation, but now \xmm\ 
interestingly fills the gaps in the $Swift/XRT$ observations
and  a flare which $Swift/XRT$ missed was caught. 
At least two source states
are observed with \xmm\, with a ``bright'' flare (lasting $\sim$1 hour) 
which reaches 
a flux at least one order of magnitude larger than the fainter state.
Intense flaring activity is observed not only in the bright state, but also in
the fainter state, with flares typically lasting few thousands seconds.

The spin period found with $RXTE$ (\citealt{Smith2006b}, \citealt{Swank2007}) 
is confirmed and for the first time
a pulse profile below 2~keV is reported (see Fig.~\ref{fig:efold}).
The hardness ratio is modulated on the pulsar spin period as well.
The pulse profiles are different in the two source states, with a double-peaked
shape during the bright phase, and a broad single pulse profile during the
faint state.

The phase resolved spectra show evidence for a spectral change during the bright flare,
where the spectrum is softer during spin-phase range [0.4--0.8], covering the secondary
minimum and the rising to the secondary maximum.

We adopted two pheonomenological models to fit the X--ray continua: 
a cut-off power-law, which
is a standard model for this kind of sources (e.g. \citealt{Coburn2002}) and a 
combination of a power-law plus a blackbody component (with kT$\sim$1--2~keV), 
also used many times in describing
the accreting pulsars X--ray emission (e.g. \citealt{Lapalombara2006}).
The resulting blackbody radii (few hundred meters) 
are consistent with emission in the accretion columns, as proposed and studied in detail
by \citet{Becker2005}.
We note that we successfully fitted the $Swift/XRT$ spectra (Paper~II) with 
the models reported here, with similar results. 
The smaller statistics (and the less extended energy range) of $Swift/XRT$ data
compared with \xmm, did not require 
models more complex than an absorbed powerlaw.

The X--ray observations of the \src\ outbursts can be used to test the different
models proposed to explain the SFXTs short outbursts.
\citet{Negueruela2005a} suggested that SFXTs are HMXBs 
in wide eccentric binaries,
in order to explain the low quiescent luminosity 
($\sim$10$^{32}$~erg~s$^{-1}$) and the
high dynamic range observed 
(L$_{\rm outburst}$/L$_{\rm quiescent}$=10$^{3}$--10$^{4}$).

	\begin{figure}[th!]
	 	\hspace{.4cm}
                 \includegraphics[angle=0,width=8cm]{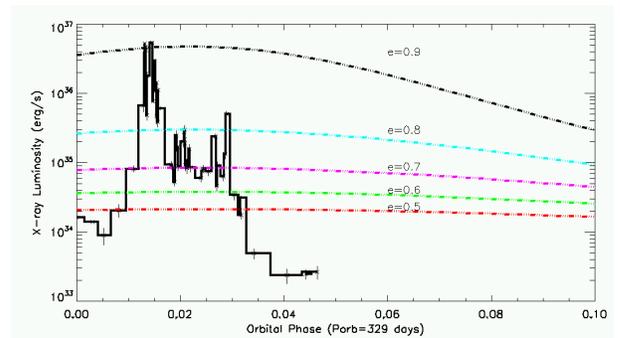}
 		\caption{Comparison of the \src\ lightcurve observed with $Swift/XRT$ with the 
          X--ray emission expected from Bondi-Hoyle accretion from the spherically symmetric
wind of the B1-type supergiant
companion, for different binary system eccentricities (see text). An orbital
period of 329~days is assumed.
		}
                \label{fig:ecc}
	\end{figure}

Several authors applied the Bondi-Hoyle accretion to wind-fed X--ray binaries in
eccentric orbits (e.g. \citealt{Raguzova1998} and references therein). 
In order to derive the X--ray luminosity variations expected from the motion
of the neutron star along the orbit around a supergiant companion, we  assume the equations
outlined in \citet{Reig2003}. 
In Fig.~\ref{fig:ecc} we compare the $Swift/XRT$ lightcurve with the X--ray emission expected
from Bondi-Hoyle accretion onto the neutron star 
from a spherically simmetric homogeneous wind from  a B1-type supergiant in an eccentric orbit.
We assumed a beta-law for the supergiant wind, with an exponent $\beta$=1,
a stellar mass of 39~M$_\odot$, a radius of 42~R$_\odot$, 
a wind terminal velocity of 1200~km~s$^{-1}$ \citep{Lefever2007}
and a wind mass loss of 3.7$\times$10$^{-6}$ ~M$_\odot$~yrs$^{-1}$ \citep{Vink2000}.
Note however that  mass, radius, wind mass loss and
terminal velocity of the supergiant donor are largely uncertain.
In any case, for any reasonable choice of the companion wind parameters, the \src\ lightcurve 
is too narrow and steep to be explained with accretion from a spherically symmetric wind,
even adopting extreme eccentricities for the binary system.

The difficulties of such simple models to explain the
short duration outbursts in some Be X--ray transients were already noticed 
(e.g. \citealt{Stella1986} or \citealt{Raguzova1998}). 
Centrifugal inhibition of the accretion, occurring when the 
magnetospheric radius becomes larger than the corotation radius, due to changes
in the wind parameters along the eccentric orbit, has been invoked as a possible
mechanism at work in these systems.
For a slow pulsator like \src\ and the wind parameters quoted above, 
the centrifugal inhibition near the periastron passage would require 
an unusual magnetic field of the order of 10$^{14}$~G.
However, the spectral cut-off at 15~keV  (Paper~I) is fully consistent with 
B$\sim$10$^{12}$~G \citep{Coburn2002}, thus leaving ample room 
for accretion to never be inhibited.

An alternative viable explanation for the sharpness of the observed X--ray lightcurve
could be that in \src\ the supergiant wind is not spherically symmetric.
We propose that the supergiant wind has a second  component (besides the ``polar''
spherically simmetric one), in the form of an ``equatorial disk'', inclined with respect to
the orbital plane by an angle $\theta$ 
(with $\theta$=90$\degmark$ if the disk is perpendicular to the orbital plane). 
The short outburst  is then produced when the neutron star crosses this equatorial disk component, 
denser and slower than the ``polar'' wind component.
Deviations from spherical symmetry in hot massive star winds
are also suggested by optical observations (see e.g. \citealt{Prinja1990}, \citealt{Prinja2002}) 
and the presence of 
equatorial disk components, denser and slower with respect to the polar wind,
also results from simulations  \citep{Ud-doula2006}.

The thickness {\em ``h''} of the  densest part of this disk originating 
from  the supergiant equatorial region
can be calculated from the duration of the brightest part of the outburst (which
lasts less than 1~day, time needed for the neutron star to cross it) and the
neutron star velocity, 100--200~km~s$^{-1}$: 
{\em h}$\sim$(0.8--1.7)$\times$sin$(\theta)$$\times$10$^{12}$~cm.

	\begin{figure}[t]
	 	\hspace{0.cm}
                 \includegraphics[angle=0,width=9.5cm]{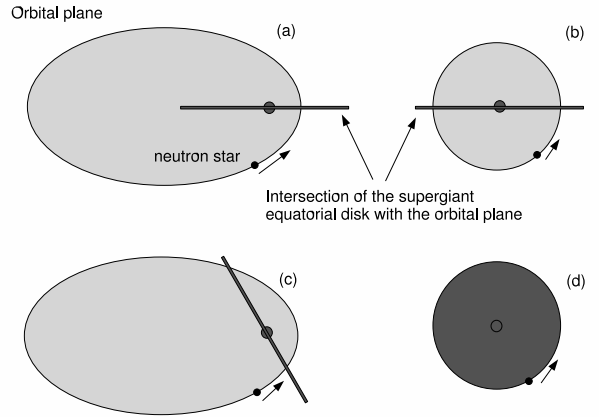}
 		\caption{Proposed geometry for different kinds of systems 
with supergiant companions. 
The ellipses (or circles, light color) mark the orbital plane. 
The black small circle indicates the neutron
star, moving along the orbit.
The  equatorial wind (in the form of a disk) from the supergiant 
is inclined with respect to the orbital plane. 
The dark line sketched here, centered on the supergiant star,
marks the intersection of the equatorial wind  with the orbital plane. 
X--ray outbursts occur when the neutron star crosses the inclined equatorial disk from the companion.
In case ``a'' the outburts are periodic
and regularly spaced, and a certain eccentricity is required in order to allow a low 
quiescent luminosity, which for a circular orbit under the influence 
of a polar wind would not be reached.
Case ``b'' is also a possibility (circular orbit with inclined equatorial disk and two outbursts
per orbit, reaching the same peak flux). 
Case ``c'' can be a possibility for all other SFXTs where a clear periodicity in the outbursts
has not been found yet. 
Case ``d'' sketches the situation where the equatorial disk lies on the orbital plane (this
is not the case for SFXTs, but maybe the case for persistent HMXBs, like Vela X-1).
		}
                \label{fig:geom}
	\end{figure}

\begin{figure*}
\vbox{
\includegraphics[height=6.5cm,angle=0]{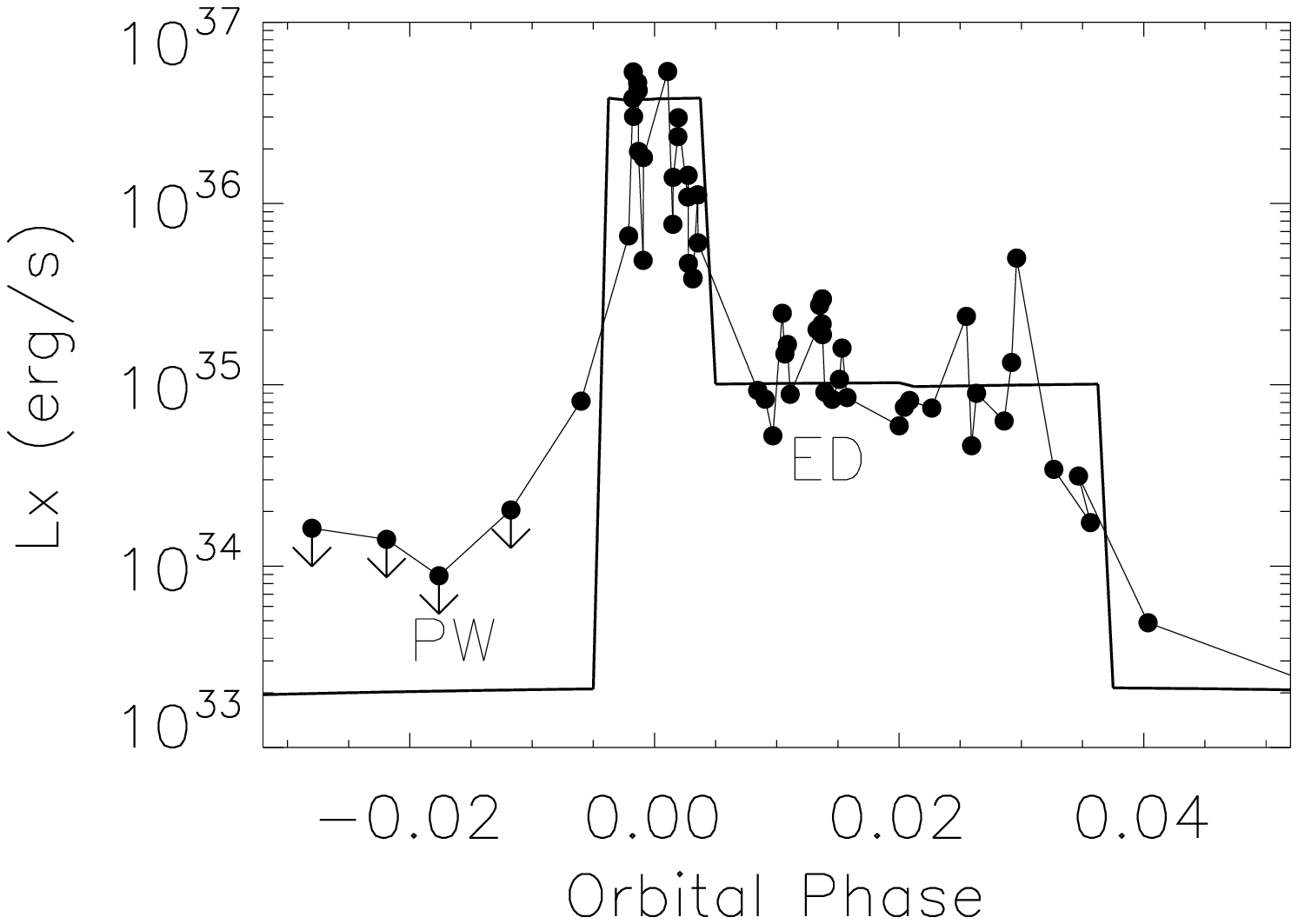}
\includegraphics[height=6.5cm,angle=0]{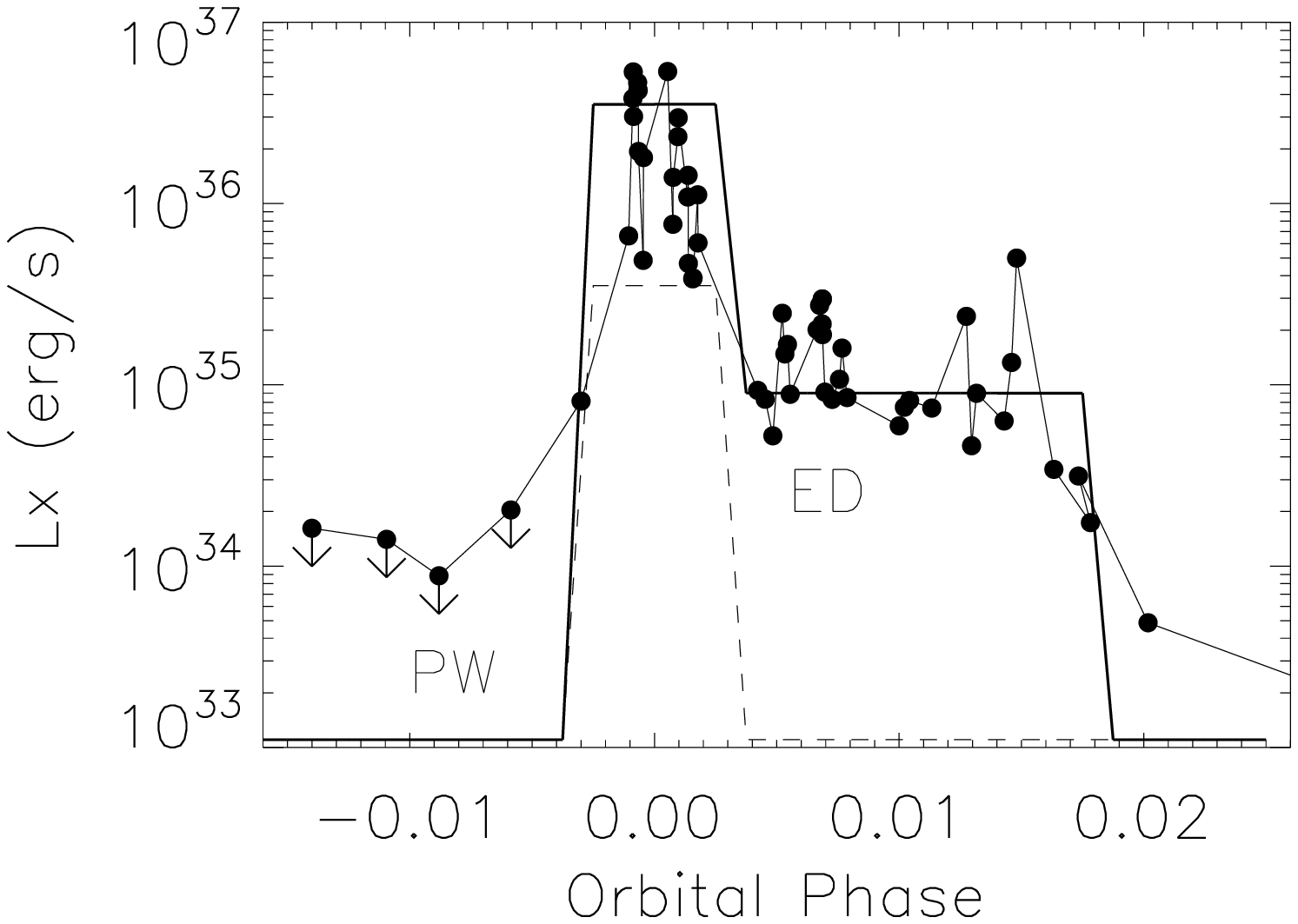}
\includegraphics[height=6.5cm,angle=0]{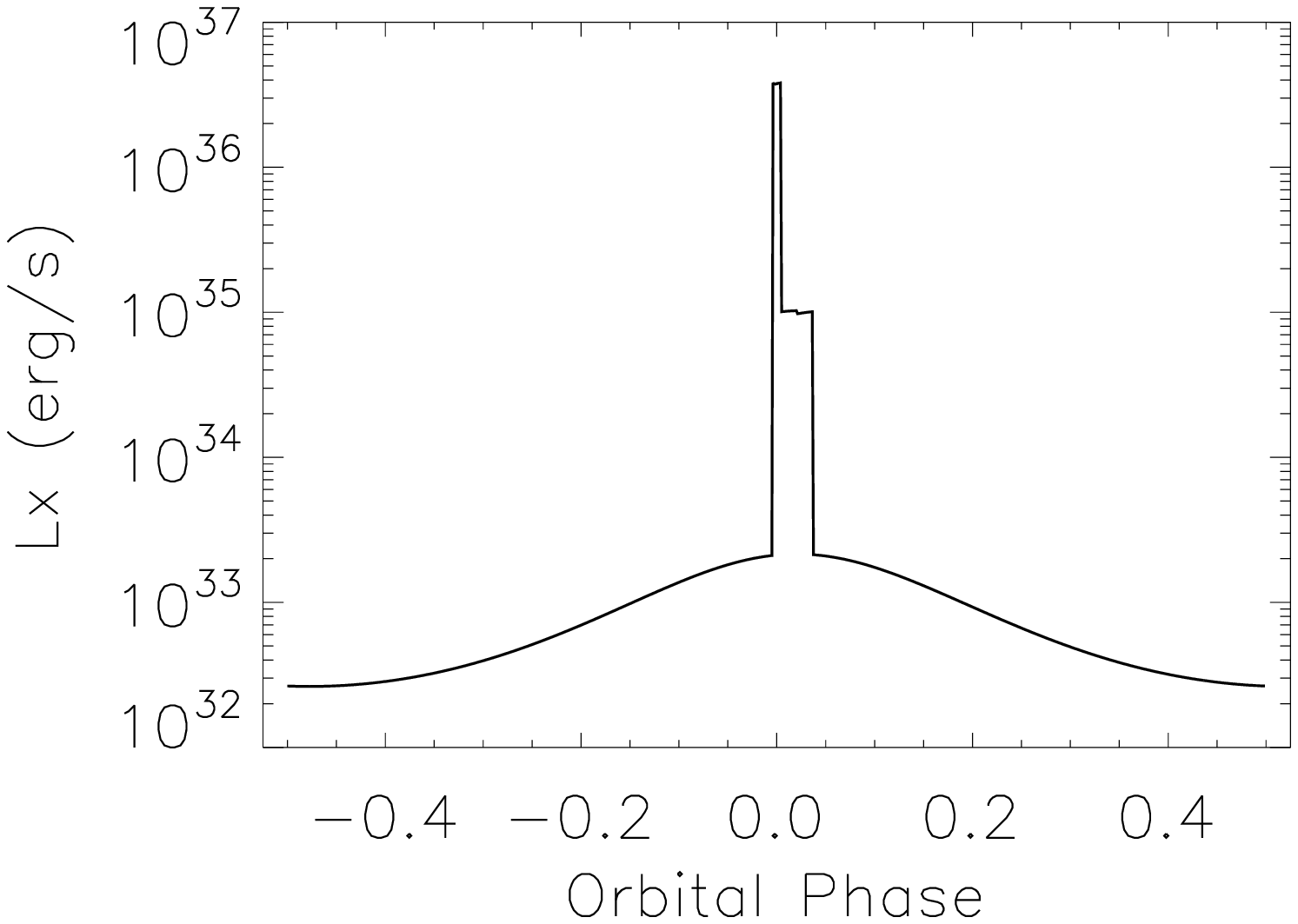}
\includegraphics[height=6.5cm,angle=0]{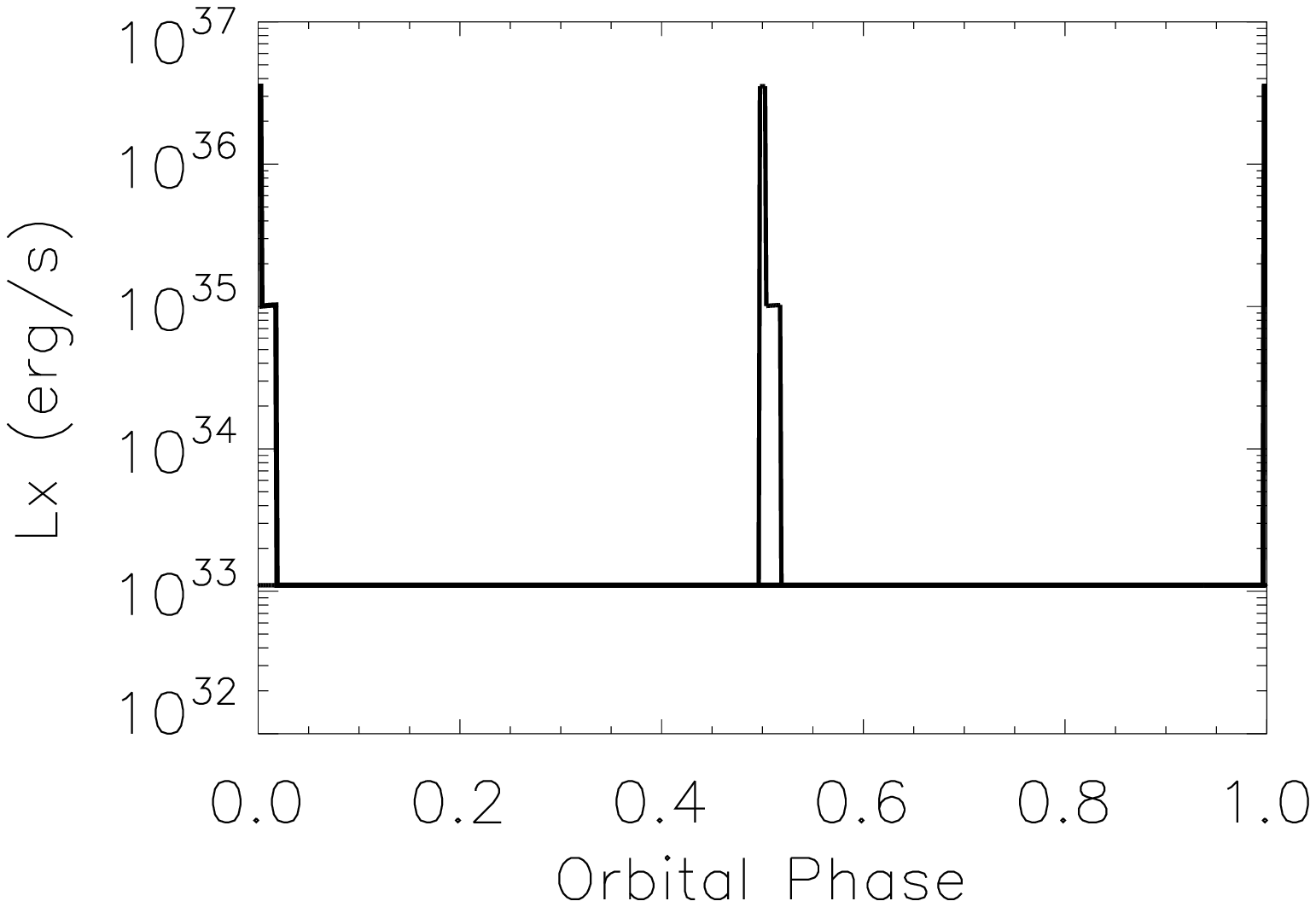}
\includegraphics[height=6.5cm,angle=0]{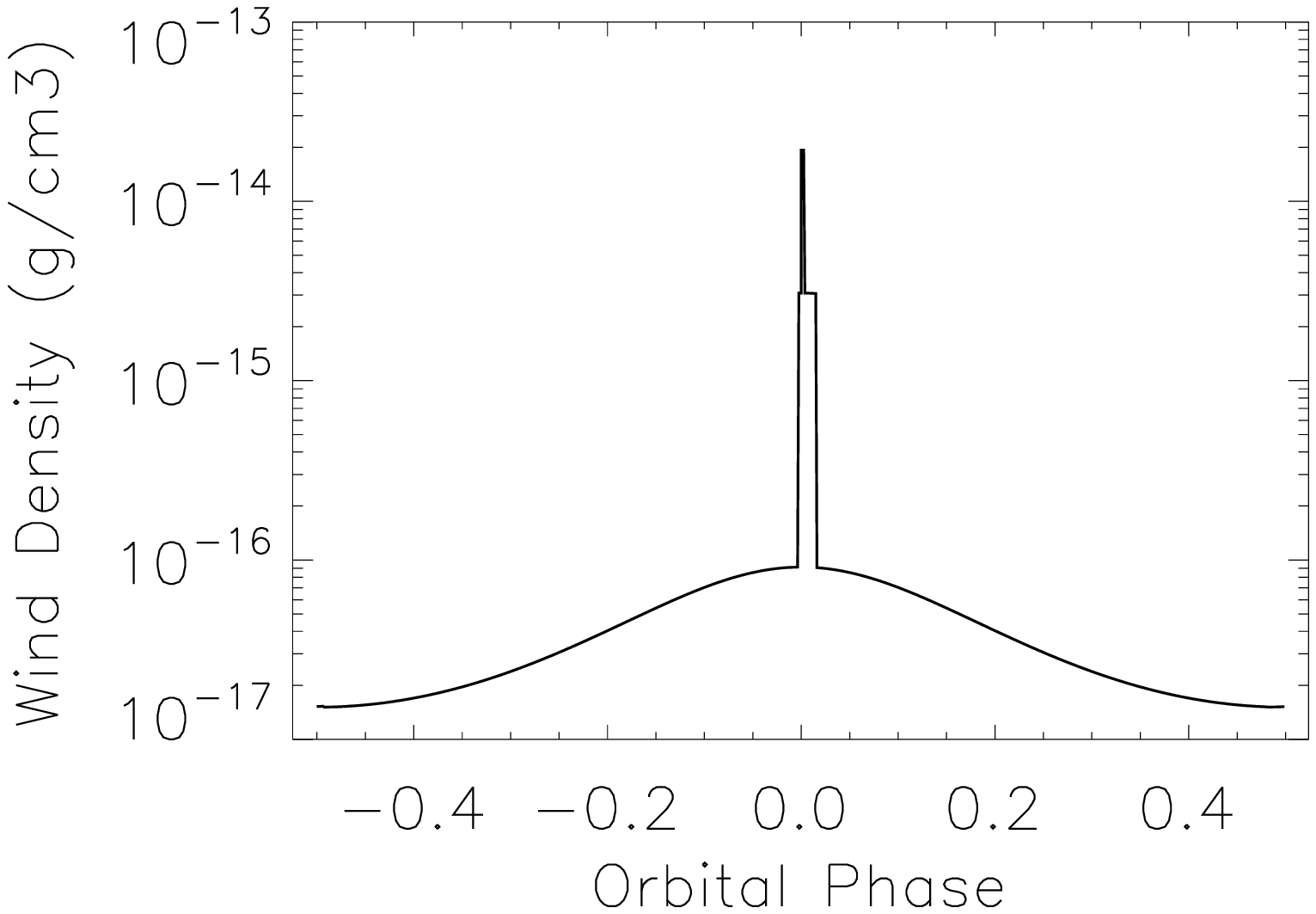}
\includegraphics[height=6.5cm,angle=0]{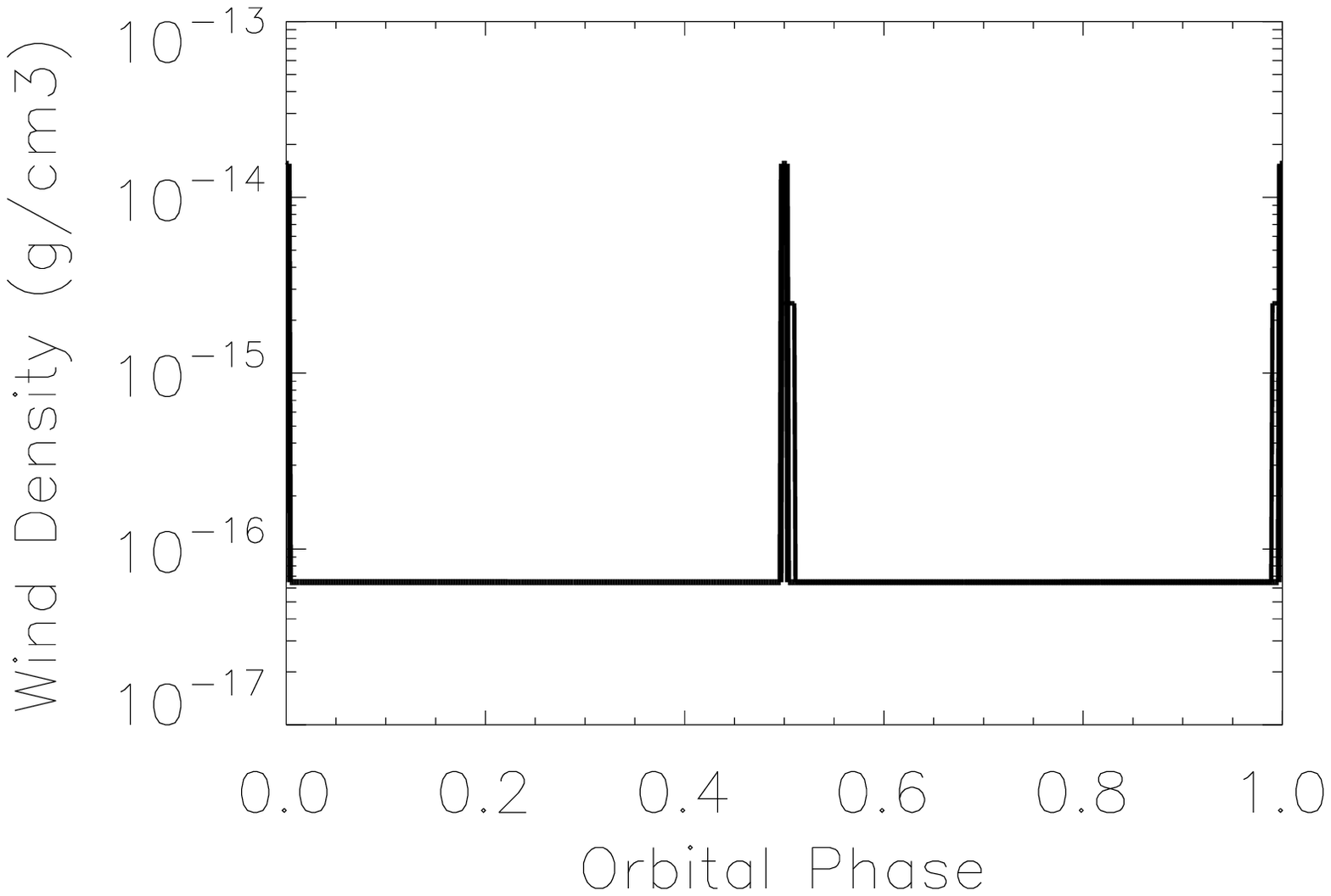}}
\caption[]{Results of the proposed model compared with the $Swift/XRT$ lightcurve ({\em upper panels})
the expected luminosity variations ({\em middle panels}) 
and wind density  along the whole  neutron star orbit 
({\em lower panels}) in two cases: {\em on the left} geometry case ``a'' (with orbital period of 164.5~days
and an assumed eccentricity of 0.4) is shown (see Fig.~\ref{fig:geom}), 
while {\em on the right} the results for case ``b'' are reported 
(orbital period of 329~days, circular orbit and two similar
outbursts per orbit). 
Both models assume a blue supergiant with a mass of  
39~M$_\odot$ and radius of 42~R$_\odot$, a  ``polar wind'' component \textbf{(``PW'')} with
a terminal velocity of 1800~km~s$^{-1}$.
The X--ray lightcurve observed with $Swift/XRT$ is better reproduced assuming
a ``polar wind'' mass loss rate of 
5$\times$10$^{-6}$~M$_\odot$~yrs$^{-1}$ for case ``a'' and
9$\times$10$^{-7}$ ~M$_\odot$~yrs$^{-1}$ for case ``b''. 
The second wind component, in form of an equatorial disk (``ED''), has a 
variable velocity ranging from 750~km~s$^{-1}$ to 1400~km~s$^{-1}$ (for case ``a''),
and from 850~km~s$^{-1}$ to 1600~km~s$^{-1}$ for case ``b''. The wind density profiles
assumed here are reported in the two lower panels.
For a magnetic field of 10$^{12}$~G the centrifugal barrier is open along all the orbit in both cases.
Dashed line in the first upper right figure shows the model prediction based on a ten times smaller wind density 
with respect to that assumed to reproduce the peak luminosity (maintaining fixed all the other parameters).
}
\label{fig:model}
\end{figure*}

The proposed geometry for the source is sketched in Fig.~\ref{fig:geom}, where four
different possibilities are outlined.
Case ``a'' represents a neutron star in an eccentric orbit around the supergiant star
with an inclined equatorial disk:
this geometry explains the periodic equally spaced outbursts and 
assumes a certain eccentricity (which is allowed  not 
to be extreme in our model,
since the outbursts are mainly produced by the geometry, and not by the high eccentricity).
On the other hand, a certain eccentricity should be present ($\sim$0.4) in order to reach 
quiescent X--ray luminosities of the order of $\sim$10$^{32}$~erg~s$^{-1}$, which would
not be allowed by the accretion from the polar component of the wind along a circular orbit.
We note that it is also possible that the quiescent level from \src\ is much higher than previously
observed for other SFXTs
(detections during quiescence are still lacking). 
If this case is the correct geometry for \src, 
the true orbital period is $\sim$165~days (see also middle panel in Fig\ref{fig:model}, where an orbital
period of 164.5~days has been assumed).
If the orbit is circular (case ``b''),
then 329~days would be the true orbital period and 
a quiescent level of $\sim$10$^{33}$~erg~s$^{-1}$
would be reached. Note however that a lower quiescent emission could be reached also in this e=0 case,
assuming a much lower mass loss rate from the polar wind component in the B-supergiant.
Case ``c'' in Fig.~\ref{fig:geom} is a possibility to explain the short outbursts from all the other
SFXTs, where a clear periodicity in the outbursts recurrence
has not yet been found. 
Case ``d''  happens when the wind from the equatorial region of the
supergiant lies in the orbital plane. 
This could be the case of the persistent  HMXBs (like Vela X--1).
\citet{Kaper1998} found that the observed X--ray luminosities in HMXBs is
about one order of magnitude higher than predicted
by Bondi-Hoyle accretion assuming standard wind mass loss rates and wind velocities. 
We suggest
that the lower wind velocity and higher density in the form of the proposed equatorial wind
could play a role in the observed  luminosity, which is higher than predicted.

An accretion disk around the neutron star is unlikely to  form, because
the wind from the supergiant  is too fast. 
According to \citet{Wang1981} and \citet{Waters1989} the relative velocity of the 
neutron star with respect to the wind should be less than about 130~km~s$^{-1}$ to allow
the formation of an accretion disk around the compact object (assuming a neutron star 
magnetic field of order of $\sim$10$^{12}$~G).
This also excludes
the possibility that the fainter X--ray emission in the \src\ lightcurve
observed by $Swift/XRT$ after February 9, is due to the X--ray emission produced from the exhaustion
of the  matter stored in the accretion disk.
In order to explain this lower level of X--ray activity, we propose a less dense region
of the equatorial disk, or a variable density inside the disk. 

In Fig.~\ref{fig:model} we show the predictions of our model, 
compared with the source X--ray 
lightcurve, with two different set of orbit and wind parameters, depending
on the orbital period assumed for the source. 
If the orbital period is 329~days, the fact that the two  consecutive outbursts
observed with \sw\ reach roughly the same flux at maximum 
(note however that the data present several gaps 
so that the maximum  might have been missed) implies a circular or a very 
low eccentricity orbit (as in case ``b'' in Fig.~\ref{fig:geom}).
This case is shown in the right panels in Fig.~\ref{fig:model}. 
If instead the  period is $\sim$164.5 days, then the outbursts 
can be explained with the accretion
from the supergiant inclined equatorial disk during the periastron passage, in an eccentric orbit 
(as in case ``a'' in Fig.~\ref{fig:geom}).  

In the caption of Fig.~\ref{fig:model} we report a possible choice for 
the  wind parameters that match the observed X--ray lightcurve.
The wind density profile is roughly in agreement with the simulations reported in \citet{Ud-doula2006}. 
For comparison, 
the winds in the  circumstellar disks around main sequence Be stars are much
denser (10$^{-12}$-10$^{-13}$~g~cm$^{-3}$, e.g. \citealt{Rinehart1999}).
We assume a two-component  wind from the companion, 
with a ``polar'' spherically simmetric wind (``PW'' in Fig.~\ref{fig:model}, 
with standard
parameters for a B-type supergiant, e.g. \citealt{Vink2000}, \citealt{Lefever2007}) 
together with  
an inclined equatorial disk (``ED'' in Fig.~\ref{fig:model}) 
with variable parameters 
in order to reproduce the variable X--ray emission.
Note however that, since the X--ray luminosity produced by the wind accretion 
is proportional to $\mdot$$v_{rel}$$^{-4}$ (where $\mdot$ is the wind mass loss rate,
and $v_{rel}$ is the relative velocity of the wind material with respect to the neutron
star, \citealt{Waters1989}), different combinations of wind density and velocity
in the equatorial disk are able to reproduce the X--ray lightcurve as well.
Thus, the wind parameters reported in Fig.~\ref{fig:model} are only an example.

The high variability with several flares observed both with \xmm\ and $Swift/XRT$ can
be explained with the clumpy nature of the wind in the equatorial disk (as originally proposed
by \citet{zand2005} to explain the outbursts from SFXTs).

The idea that the neutron star intersects an equatorial 
wind component once or twice during its
orbit  to explain  recurrent flares,  is not new in the study of Be/X--ray binaries 
(see, e.g., the Be-transient A0538--66, \citealt{Apparao1985}). 
In this respect, the fact that the spin and orbital periods
($\sim$329~days or, more likely, a half of this) place \src\ inside the locus typical for 
Be/X--ray transients of the Corbet diagram \citep{Corbet1986} 
is puzzling, since the \src\ donor star is a confirmed
B-type supergiant, and not a Be star. 
Moreover, the source outburst duration  appears significantly
shorter than that typically displayed by Be/X--ray binaries.
On the other hand, there are other exceptions: for example, 
the X--ray binary SAX~J2103.5+4545, optically identified
with a Be star, lies in the region of the Corbet diagram proper for
supergiant wind-fed HMXBs \citep{Baykal2007}.

\section{Conclusions}
\label{conclusions}

We have  proposed here a new explanation for the short outbursts in SFXTs, based on 
our results of a monitoring campain of \src. 
An equatorial wind from the  supergiant companion is suggested, based
on the narrow and steep shape of the X--ray lightcurve observed during the latest outburst.
The short outburst is suggestive of a  deviation from coplanarity ($\theta$$>$0) of the equatorial
plane of the companion with the orbital plane (as, e.g., in PSR~B1259-63, \citealt{Wex1998}
or in PSR~J0045--7319, \citealt{Kaspi1994}), and of a some degree of eccentricity (e$>$0).

Both orbital eccentricity and no-coplanarity can be explained 
by a substantial supernova ``kick'' (e.g. \citealt{Eggleton2001})
at birth. This could suggest that SFXTs are likely young systems, probably younger than
persistent HMXBs. 

We are aware that the derived quantities for the proposed supergiant equatorial disk 
are highly uncertain and still speculative. 
This uncertainty clearly emerges from the fact that 
the observed X--ray lightcurve can be reproduced
with both orbital periods (329~days and 164.5~days), 
different eccentricities  and different wind properties.
A determination of the orbital eccentricity of the system, for example, would be  essential in
better constraining the expected properties of the wind which
match with the level of X--ray emission during the outbursts. 
These considerations highlight the need for an as complete as possible
deep monitoring of the X--ray emission along the orbit, 
together with optical observations in order to
constrain the supergiant wind properties.

\onltab{2}{        
 \begin{table*}
 \begin{center}
 \caption{Observation log.}
 \label{igr112:tab:alldata2}
 \begin{tabular}{lllll}
 \hline
 \hline
 \noalign{\smallskip}
Sequence         & Start time (MJD)     & Start time  (UT)             & End time   (UT)                 & Net Exposure$^{\mathrm{a}}$    \\
                 &                      & (yyyy-mm-dd hh:mm:ss)           & (yyyy-mm-dd hh:mm:ss)        &(s)       \\
 \noalign{\smallskip}
 \hline
 \noalign{\smallskip}
00030881024     & 54256.7449	  &	  2007-06-05 17:52:35	  &	  2007-06-05 21:15:58	  &	  1519    \\
00030881025     & 54263.4918	  &	  2007-06-12 11:48:11	  &	  2007-06-12 16:57:58	  &	  1522    \\
00030881026     & 54270.3808	  &	  2007-06-19 09:08:17	  &	  2007-06-19 11:05:57	  &	  1843    \\
00030881027     & 54277.1485	  &	  2007-06-26 03:33:52	  &	  2007-06-26 05:27:57	  &	  2109    \\
00030881028     & 54284.8571	  &	  2007-07-03 20:34:14	  &	  2007-07-03 23:59:57	  &	  2347    \\
00030969001     & 54294.5627	  &	  2007-07-13 13:30:15	  &	  2007-07-14 07:10:56	  &	  2126    \\
00030881030     & 54298.0425	  &	  2007-07-17 01:01:12	  &	  2007-07-17 02:58:57	  &	  2069    \\
00030881031     & 54305.0012   	  &       2007-07-24 00:01:42     &       2007-07-24 13:16:55     &       1551    \\
00030881032     & 54312.2343      &       2007-07-31 05:37:23     &       2007-07-31 10:26:56     &       1695    \\
  \noalign{\smallskip}
  \hline
  \end{tabular}
  \end{center}
  \begin{list}{}{}
  \item[$^{\mathrm{a}}$] The exposure time is spread over several snapshots (single continuous pointings at the target)
        during each observation.
  \end{list}
  \end{table*}
} 
%


\begin{acknowledgements}
Based on observations obtained with XMM-Newton, an ESA science
mission with instruments and contributions directly funded by ESA
member states and the USA (NASA).
Based on observations with INTEGRAL, an ESA project with instruments and the science data 
centre funded by ESA member states (especially the PI countries: 
Denmark, France, Germany, Italy, Switzerland, Spain), Czech Republic and Poland, 
and with the participation of Russia and the USA.
We thank the \xmm, \inte, and $Swift$ teams for making these observations possible,
in particular the duty scientists and science planners.
LS would like to thank all the partecipants
to a seminar at INAF/IASF-Rome, on 2nd March, 2007, which raised very interesting discussions
about the source nature, in particular  A.\ Bazzano, P.\ Persi,  V.F.\ Polcaro and P.\ Ubertini.
PR thanks INAF-IASFMi for their kind hospitality. DG
acknowledges the French Space Agency (CNES) for financial support.
This work was supported by  contract ASI/INAF I/023/05/0. 
\end{acknowledgements}

\bibliographystyle{aa} 
\bibliography{biblio}

\begin{thebibliography}{34}
\expandafter\ifx\csname natexlab\endcsname\relax\def\natexlab#1{#1}\fi

\bibitem[{{Apparao}(1985)}]{Apparao1985}
{Apparao}, K.~M.~V. 1985, \apj, 292, 257

\bibitem[{{Baykal} {et~al.}(2007){Baykal}, {Inam}, {Stark}, {Heffner},
  {Erkoca}, \& {Swank}}]{Baykal2007}
{Baykal}, A., {Inam}, S.~{\c C}., {Stark}, M.~J., {et~al.} 2007, \mnras, 374,
  1108

\bibitem[{{Becker} \& {Wolff}(2005)}]{Becker2005}
{Becker}, P.~A. \& {Wolff}, M.~T. 2005, \apj, 630, 465

\bibitem[{{Coburn} {et~al.}(2002){Coburn}, {Heindl}, {Rothschild}, {Gruber},
  {Kreykenbohm}, {Wilms}, {Kretschmar}, \& {Staubert}}]{Coburn2002}
{Coburn}, W., {Heindl}, W.~A., {Rothschild}, R.~E., {et~al.} 2002, \apj, 580,
  394

\bibitem[{{Corbet}(1986)}]{Corbet1986}
{Corbet}, R.~H.~D. 1986, \mnras, 220, 1047

\bibitem[{{Eggleton} \& {Kiseleva-Eggleton}(2001)}]{Eggleton2001}
{Eggleton}, P.~P. \& {Kiseleva-Eggleton}, L. 2001, \apj, 562, 1012

\bibitem[{{in't Zand}(2005)}]{zand2005}
{in't Zand}, J.~J.~M. 2005, \aap, 441, L1

\bibitem[{{Kaper}(1998)}]{Kaper1998}
{Kaper}, L. 1998, in Astronomical Society of the Pacific Conference Series,
  Vol. 131, Properties of Hot Luminous Stars, ed. I.~{Howarth}, 427

\bibitem[{{Kaspi} {et~al.}(1994){Kaspi}, {Johnston}, {Bell}, {Manchester},
  {Bailes}, {Bessell}, {Lyne}, \& {D'Amico}}]{Kaspi1994}
{Kaspi}, V.~M., {Johnston}, S., {Bell}, J.~F., {et~al.} 1994, \apjl, 423, L43

\bibitem[{{La Palombara} \& {Mereghetti}(2006)}]{Lapalombara2006}
{La Palombara}, N. \& {Mereghetti}, S. 2006, \aap, 455, 283

\bibitem[{{Lefever} {et~al.}(2007){Lefever}, {Puls}, \& {Aerts}}]{Lefever2007}
{Lefever}, K., {Puls}, J., \& {Aerts}, C. 2007, \aap, 463, 1093

\bibitem[{{Lubinski} {et~al.}(2005){Lubinski}, {Bel}, {von Kienlin},
  {Budtz-Jorgensen}, {McBreen}, {Kretschmar}, {Hermsen}, \&
  {Shtykovsky}}]{Lubinski2005}
{Lubinski}, P., {Bel}, M.~G., {von Kienlin}, A., {et~al.} 2005, The
  Astronomer's Telegram, 469

\bibitem[{{Masetti} {et~al.}(2006){Masetti}, {Pretorius}, {Palazzi}, {Bassani},
  {Bazzano}, {Bird}, {Charles}, {Dean}, {Malizia}, {Nkundabakura}, {Stephen},
  \& {Ubertini}}]{Masetti2006}
{Masetti}, N., {Pretorius}, M.~L., {Palazzi}, E., {et~al.} 2006, \aap, 449,
  1139

\bibitem[{{Negueruela} {et~al.}(2005){Negueruela}, {Smith}, \&
  {Chaty}}]{Negueruela2005b}
{Negueruela}, I., {Smith}, D.~M., \& {Chaty}, S. 2005, The Astronomer's
  Telegram, 470

\bibitem[{{Negueruela} {et~al.}(2006){Negueruela}, {Smith}, {Reig}, {Chaty}, \&
  {Torrej{\'o}n}}]{Negueruela2005a}
{Negueruela}, I., {Smith}, D.~M., {Reig}, P., {Chaty}, S., \& {Torrej{\'o}n},
  J.~M. 2006, in Proceedings of the ''The X-ray Universe 2005'', 26-30
  September 2005, El Escorial, Madrid, Spain. Ed. by A. Wilson. ESA SP-604,
  Volume 1, Noordwijk: ESA Publications Division, ISBN 92-9092-915-4, 2006, ed.
  A.~{Wilson}, 165

\bibitem[{{Prinja}(1990)}]{Prinja1990}
{Prinja}, R.~K. 1990, \aap, 232, 119

\bibitem[{{Prinja} {et~al.}(2002){Prinja}, {Massa}, \&
  {Fullerton}}]{Prinja2002}
{Prinja}, R.~K., {Massa}, D., \& {Fullerton}, A.~W. 2002, \aap, 388, 587

\bibitem[{{Raguzova} \& {Lipunov}(1998)}]{Raguzova1998}
{Raguzova}, N.~V. \& {Lipunov}, V.~M. 1998, \aap, 340, 85

\bibitem[{{Reig} {et~al.}(2003){Reig}, {Rib{\'o}}, {Paredes}, \&
  {Mart{\'{\i}}}}]{Reig2003}
{Reig}, P., {Rib{\'o}}, M., {Paredes}, J.~M., \& {Mart{\'{\i}}}, J. 2003, \aap,
  405, 285

\bibitem[{{Rinehart} {et~al.}(1999){Rinehart}, {Houck}, \&
  {Smith}}]{Rinehart1999}
{Rinehart}, S.~A., {Houck}, J.~R., \& {Smith}, J.~D. 1999, \aj, 118, 2974

\bibitem[{{Romano} {et~al.}(2007){Romano}, {Sidoli}, {Mangano}, {Mereghetti},
  \& {Cusumano}}]{Romano2007}
{Romano}, P., {Sidoli}, L., {Mangano}, V., {Mereghetti}, S., \& {Cusumano}, G.
  2007, \aap, 469, L5

\bibitem[{{Sguera} {et~al.}(2005){Sguera}, {Barlow}, {Bird}, {Clark}, {Dean},
  {Hill}, {Moran}, {Shaw}, {Willis}, {Bazzano}, {Ubertini}, \&
  {Malizia}}]{Sguera2005}
{Sguera}, V., {Barlow}, E.~J., {Bird}, A.~J., {et~al.} 2005, \aap, 444, 221

\bibitem[{{Sidoli} {et~al.}(2006){Sidoli}, {Paizis}, \&
  {Mereghetti}}]{SidoliPM2006}
{Sidoli}, L., {Paizis}, A., \& {Mereghetti}, S. 2006, \aap, 450, L9

\bibitem[{{Smith} {et~al.}(2006{\natexlab{a}}){Smith}, {Bezayiff}, \&
  {Negueruela}}]{Smith2006b}
{Smith}, D.~M., {Bezayiff}, N., \& {Negueruela}, I. 2006{\natexlab{a}}, The
  Astronomer's Telegram, 773

\bibitem[{{Smith} {et~al.}(2006{\natexlab{b}}){Smith}, {Bezayiff}, \&
  {Negueruela}}]{Smith2006a}
{Smith}, D.~M., {Bezayiff}, N., \& {Negueruela}, I. 2006{\natexlab{b}}, The
  Astronomer's Telegram, 766

\bibitem[{{Steeghs} {et~al.}(2006){Steeghs}, {Torres}, \&
  {Jonker}}]{Steeghs2006}
{Steeghs}, D., {Torres}, M.~A.~P., \& {Jonker}, P.~G. 2006, The Astronomer's
  Telegram, 768

\bibitem[{{Stella} {et~al.}(1986){Stella}, {White}, \& {Rosner}}]{Stella1986}
{Stella}, L., {White}, N.~E., \& {Rosner}, R. 1986, \apj, 308, 669

\bibitem[{{Swank} {et~al.}(2007){Swank}, {Smith}, \& {Markwardt}}]{Swank2007}
{Swank}, J.~H., {Smith}, D.~M., \& {Markwardt}, C.~B. 2007, The Astronomer's
  Telegram, 999

\bibitem[{{ud-Doula} {et~al.}(2006){ud-Doula}, {Townsend}, \&
  {Owocki}}]{Ud-doula2006}
{ud-Doula}, A., {Townsend}, R.~H.~D., \& {Owocki}, S.~P. 2006, \apjl, 640, L191

\bibitem[{{Vink} {et~al.}(2000){Vink}, {de Koter}, \& {Lamers}}]{Vink2000}
{Vink}, J.~S., {de Koter}, A., \& {Lamers}, H.~J.~G.~L.~M. 2000, \aap, 362, 295

\bibitem[{{Walter} {et~al.}(2003){Walter}, {Rodriguez}, {Foschini}, {de Plaa},
  {Corbel}, {Courvoisier}, {den Hartog}, {Lebrun}, {Parmar}, {Tomsick}, \&
  {Ubertini}}]{Walter2003}
{Walter}, R., {Rodriguez}, J., {Foschini}, L., {et~al.} 2003, \aap, 411, L427

\bibitem[{{Wang}(1981)}]{Wang1981}
{Wang}, Y.-M. 1981, \aap, 102, 36

\bibitem[{{Waters} {et~al.}(1989){Waters}, {de Martino}, {Habets}, \&
  {Taylor}}]{Waters1989}
{Waters}, L.~B.~F.~M., {de Martino}, D., {Habets}, G.~M.~H.~J., \& {Taylor},
  A.~R. 1989, \aap, 223, 207

\bibitem[{{Wex} {et~al.}(1998){Wex}, {Johnston}, {Manchester}, {Lyne},
  {Stappers}, \& {Bailes}}]{Wex1998}
{Wex}, N., {Johnston}, S., {Manchester}, R.~N., {et~al.} 1998, \mnras, 298, 997

\end{thebibliography}

\end{document}